\documentclass[
11pt,%
tightenlines,%
twoside,%
onecolumn,%
nofloats,%
nobibnotes,%
nofootinbib,%
superscriptaddress,%
noshowpacs,%
centertags]%
{revtex4}
\usepackage{ljm}

\usepackage{booktabs}

\usepackage{braket}
\usepackage{physics}
\usepackage{algorithm}
\usepackage{algpseudocode}

\algnewcommand\And{\textbf{ and }}
\algnewcommand\Or{\textbf{ or }}
\algnewcommand\Elif{\newline\textbf{else if }}
\algnewcommand\Then{\textbf{ then}}
\usepackage{amsmath,amsfonts,bm,amssymb,mathtools}

\pdfmapfile{=literat.map}

\newcommand{\e}{\textrm{e}}
\newcommand{\ii}{\textrm{i}}

\begin{document}

\titlerunning{Single-Qubit Quantum Neural Networks}
\authorrunning{Souza et al.}

\title{Regression and Classification with Single-Qubit Quantum Neural Networks}

\author{\firstname{Leandro C.}~\surname{Souza}}
\affiliation{National Laboratory of Scientific Computing (LNCC), Petr\'opolis, RJ, 25651-075, Brazil}
\affiliation{Universidade Federal da Para\'{\i}ba (UFPB), Jo\~ao Pessoa, 58051-900, PB, Brazil}

\author{\firstname{Bruno C.}~\surname{Guingo}}
\affiliation{National Laboratory of Scientific Computing (LNCC), Petr\'opolis, RJ, 25651-075, Brazil}
\affiliation{Universidade Cat\'olica de Petr\'opolis (UCP), Petr\'opolis, 25610-130, RJ, Brazil}

\author{\firstname{Gilson}~\surname{Giraldi}}
\affiliation{National Laboratory of Scientific Computing (LNCC), Petr\'opolis, RJ, 25651-075, Brazil}

\author{\firstname{Renato}~\surname{Portugal}}
\email[E-mail: ]{portugal@lncc.br}
\affiliation{National Laboratory of Scientific Computing (LNCC), Petr\'opolis, RJ, 25651-075, Brazil}
\affiliation{Universidade Cat\'olica de Petr\'opolis (UCP), Petr\'opolis, 25610-130, RJ, Brazil}

\received{;;;}

\begin{abstract}
The literature reflects a mutually beneficial relationship between machine learning and quantum computing, where progress in one field frequently drives improvements in the other. Motivated by the rich connection between these areas, we use a resource-efficient and scalable Single-Qubit Quantum Neural Network (SQQNN) for both regression and classification tasks using a new data uploading technique. The SQQNN leverages parameterized single-qubit unitary operators and quantum measurements to achieve efficient learning. To train the model, we use gradient descent for regression tasks. For classification, we introduce a novel training method inspired by polynomial regression, which can efficiently find a global minimizer of the transformed least-squares objective in a single step. This approach significantly accelerates training compared to iterative methods. Evaluated across various applications, the SQQNN exhibits virtually error-free and strong performance in regression and classification tasks, including Wisconsin Breast Cancer and MNIST datasets. These results demonstrate the versatility, scalability, and suitability of the SQQNN for deployment on near-term quantum devices.
\end{abstract}



\maketitle

\section{Introduction}
\label{sec-intro}

Classical machine learning, especially through artificial neural networks, has transformed numerous fields by enabling pattern recognition, prediction, and complex decision-making tasks. At the core of these networks lies the neuron, a computational unit inspired by biological neurons~\cite{mcculloch1943}. One of the earliest models, the perceptron~\cite{rosenblatt1958}, acts as a binary classifier by producing an output based on a weighted sum of inputs~\cite{Goodfellow2016}. Despite its simplicity, the perceptron laid the groundwork for more sophisticated architectures that underpin support vector machines and modern deep learning~\cite{Roberts2022}.

Quantum computing introduces new possibilities for machine learning by leveraging quantum mechanics to potentially surpass classical computational limits~\cite{Biamonte2017,Liu2024,Gujju2024,Wang2024}. A key area of exploration is Quantum Neural Networks (QNNs), which use quantum parallelism and entanglement to process complex data efficiently. Many QNNs incorporate quantum versions of classical neurons, quantum perceptrons~\cite{Schuld2015,Kapoor2016,Adenilton2016,Wan2017} which, unlike their classical counterparts, rely on quantum measurement-induced probabilistic behaviors to simulate non-linear activations~\cite{Tacchino2019}. Recent advancements have extended QNNs to specialized quantum algorithms~\cite{Schuld2019,Cong2019,Henderson2020,Mangini2020,Bokhan2022,Benatti2022,Carvalho2024}, offering frameworks for supervised and unsupervised learning~\cite{Kouda2005} and sparking innovation in hardware implementations~\cite{Pechal2022}. Techniques such as quantum phase estimation and amplitude amplification further enrich the capabilities of quantum perceptrons, empowering QNNs to address data structures and tasks that often challenge classical models~\cite{Schuld2018book1,Schuld2021book2}.

Parameterized quantum circuits (PQCs) are central to quantum machine learning, providing flexible, tunable models that represent complex data through sequences of parameterized quantum gates~\cite{Sim2019,Benedetti2019,Ding2024}. Integrated into hybrid quantum-classical algorithms and trained via classical optimizers such as gradient descent, PQCs enable quantum neural networks (QNNs) to learn patterns effectively~\cite{Du2020}. By achieving high expressivity with fewer qubits, they support efficient architectures compatible with NISQ hardware~\cite{Sim2019,Schuld2021}. A minimal yet powerful realization of this idea is a QNN composed of single-qubit neurons with carefully chosen rotations, which can capture complex relationships while requiring minimal quantum resources.

Recent studies have explored this single-qubit approach for a variety of learning tasks~\cite{Karimi2023, Salinas2020, Salinas2021, Yu2022, Tapia2023, McCaldin2024, McFarthing2024, Cuellar2024}. Early works~\cite{Karimi2023} used the 1-qubit deterministic quantum computing model~\cite{Knill1998} to estimate intractable kernel functions, while Ref.~\cite{Salinas2020} introduced a classifier based on single-qubit rotations and data re-uploading. This was extended in~\cite{Salinas2021} to approximate any bounded complex function, with both theoretical and experimental support. Later studies~\cite{Yu2022, Tapia2023, McCaldin2024, Cuellar2024} demonstrated the applicability of single-qubit systems to supervised, unsupervised, and reinforcement learning, showing that such models can rival classical methods while remaining resource-efficient.

In this work, we propose a Single-Qubit Quantum Neural Network (SQQNN) composed of parameterized single-qubit neurons. This architecture generalizes the proposal of Ref.~\cite{Salinas2020} by introducing a more flexible data re-uploading strategy capable of handling multi-dimensional inputs directly, whereas the original approach decomposed them into several three-dimensional inputs, increasing circuit depth. The SQQNN provides a compact and resource-efficient framework suitable for regression and binary classification, offering advantages such as simplified implementation, faster execution, and reduced hardware demands. Its design leverages the expressiveness of parameterized quantum measurements to simulate non-linear activation functions, making it a practical and scalable alternative to multi-qubit models for near-term quantum devices.

The theoretical foundation of the SQQNN is based on general single-qubit unitary operations with adaptive input states and observables. We validate its capabilities across a variety of tasks, including logical gate learning, synthetic regressions, and real-world classification datasets such as MNIST. Two distinct training approaches are employed: gradient descent, used for logic and regression tasks, and a matrix-based method inspired by polynomial regression, which is effective for classification. The network consistently achieves strong performance with low error rates and fast training times (typically under a few seconds on a standard laptop), demonstrating its scalability and ability to learn both linear and non-linear relationships. All results were obtained using classical simulations of quantum circuits. Although the shallow depth and single-qubit nature of the circuits are favorable for near-term devices, their performance on real quantum hardware requires experimental validation. This is because single-qubit gates are among the most reliable operations on current quantum devices, exhibiting minimal decoherence and gate error compared to multi-qubit operations, which tend to suffer from significantly higher noise levels (see, for instance,~\cite{QMLP2022}). These results underscore the versatility of the SQQNN and its potential as a lightweight quantum model for diverse machine learning applications.

The structure of this paper is as follows: In Sec.~\ref{sec-perceptron}, we introduce the single-qubit quantum neuron model and describe how quantum gates are used to encode learnable parameters. In Sec.~\ref{sec-QNN}, we extend this model to define the Single-Qubit Quantum Neural Network (SQQNN), detailing its architecture, parametrization, and expressiveness. Sec.~\ref{sec-LM} presents two training strategies: gradient descent and a polynomial-based linear least squares method tailored for classification. In Sec.~\ref{sec-Regression}, we evaluate the SQQNN on regression tasks, including logic gate modeling, the sinc function, and two real-world datasets. In Sec.~\ref{sec-Class}, we assess the network's classification performance using datasets such as Wisconsin Breast Cancer and MNIST. Sec.~\ref{sec-comp} provides a comparative analysis with existing quantum and classical models, highlighting the performance of the SQQNN. In Sec.~\ref{sec-multiple}, we discuss how the proposed model can be extended to multi-qubit architectures, enabling multi-class classification with quantum circuits based on standard state preparation techniques. Finally, in Sec.~\ref{sec-conc}, we draw our conclusions.

\section{Single-Qubit Neuron Model}\label{sec-perceptron}

A quantum neuron is a quantum-inspired adaptation of the classical neuron, designed for supervised learning tasks. Given a dataset 
\begin{equation*}
\mathbb{D}=\left\{ \left({x}_{1},y_{1}\right),\left({x}_{2},y_{2}\right),\cdots,\left({x}_{n},y_{n}\right)\right\} \subset\mathbb{R}^{p}\times [-1,1],
\end{equation*} 
where each input ${x}_i \in \mathbb{R}^p$ is a feature vector and each corresponding output $y_i \in [-1,1]$ is the target of the quantum neuron. Quantum computational principles are applied to predict the output $y$ for an unseen input ${x} \in \mathbb{R}^p$. The weights of the quantum neuron are encoded within the angles of 1-qubit rotation gates 
\begin{eqnarray*}
R_x(\theta) =\,&  \left[\begin{array}{cc}
 \,\,\,\,\,\,\cos\frac{\theta}{2} &  -\ii\sin\frac{\theta}{2} \vspace{2pt}\\
 -\ii\sin\frac{\theta}{2}   &  \,\,\,\,\,\,\cos\frac{\theta}{2}
\end{array}\right],\,\,\,\,
R_y(\theta) =  \left[\begin{array}{cc}
 \cos\frac{\theta}{2} &  -\sin\frac{\theta}{2} \vspace{2pt}\\
 \sin\frac{\theta}{2}   &  \,\,\,\,\cos\frac{\theta}{2}
\end{array}\right],\,\,\,\,
R_z(\theta)  =
\left[\begin{array}{cc}
 \e^{-\ii \theta/2} &  0 \\
 0   &  \e^{ \ii \theta/2}
\end{array}\right].
\end{eqnarray*}
Those gates are combined to build a general unitary operator that depends on three real parameters. Since the most general 1-qubit unitary operator can be expressed as $\e^{\ii \delta} R_z(\gamma) R_y(\beta) R_z(\alpha)$~\cite{NC00}, we employ the most general single-qubit quantum neuron $\mathcal{N}$, modulo a global phase, as
\begin{equation*}
\mathcal{N}=R_z(\gamma) R_y(\beta) R_z(\alpha).  
\end{equation*} 

\begin{figure}[!h]
\centering 
\includegraphics[scale=0.22]{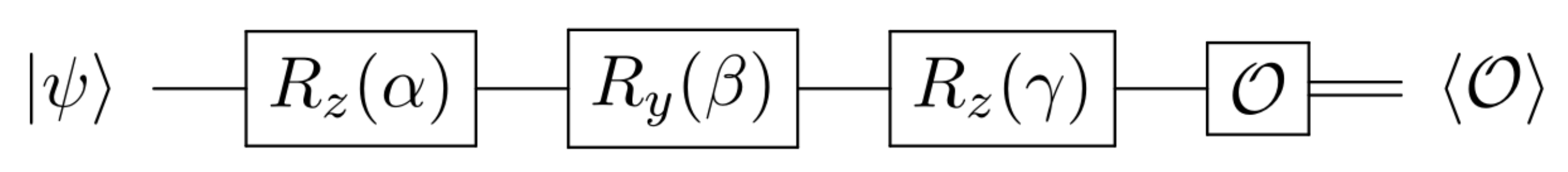}
\caption{Representation of the most general quantum neuron with an arbitrary input and measurement. The output corresponds to the average value obtained from measuring the observable $\mathcal{O}$.}\label{fig-percepton} 
\end{figure}

Fig.~\ref{fig-percepton} depicts the most general version of the single-qubit quantum neuron together with its input and a measurement. The output depends on the angles $\alpha(x)$, $\beta(x)$, and $\gamma(x)$, which are functions of the data input with a weight vector and are used in the rotation gates $R_y$ and $R_z$. The input to the circuit is the quantum state $\ket{\psi} = \cos(\theta/2)\ket{0} + \e^{\ii \phi}\sin(\theta/2)\ket{1}$, where $\theta$ and $\phi$ are two extra parameters.

After applying the rotation gates, an observable $\mathcal{O}$ is measured. The observable can be decomposed in terms of orthogonal projections as $\mathcal{O} = \lambda_0\mathcal{P}_0+\lambda_1\mathcal{P}_1$, where $\lambda_0$ and $\lambda_1$ are real numbers and the projector $\mathcal{P}_1$ is defined as
\begin{equation*}
    \mathcal{P}_1 =  \left[ \begin {array}{cc}  \sin^{2} {\frac {\omega}{2}}&{{\e}^{-\ii\varphi}}\cos  {\frac {\omega}{2}}  \sin  {\frac {\omega}{2}}  
\\ \noalign{\medskip}{{\e}^{\ii\varphi}}\cos  {\frac {\omega}{
2}}  \sin  {\frac {\omega}{2}}  &  \cos^{2}
  {\frac {\omega}{2}}    \end {array} \right],
\end{equation*}
and its orthogonal complement is $\mathcal{P}_0 = I - \mathcal{P}_1$, so that the measurement outcome is either $\lambda_0$ or $\lambda_1$. Without loss of generality, we set $\lambda_0 = +1$ and $\lambda_1 = -1$. Note that the output of the quantum neuron is not the measurement outcome but the average $ \left\langle\mathcal{O}\right\rangle$. Because quantum computers usually support only measurements in the computational basis, we implement the measurement of $\mathcal{O}$ using the unitary operator 
\begin{equation*}
    U_{\mathcal{O}}=\left[ \begin {array}{cc} \cos  {\frac {\omega}{2}}  &-{
{\e}^{-\ii\varphi}}\sin  {\frac {\omega}{2}}  
\\ \noalign{\medskip}{{\e}^{\ii\varphi}}\sin  {\frac {\omega}{2
}}  &\cos  {\frac {\omega}{2}}  \end {array}
 \right] ,
\end{equation*}
obtained from the 1-eigenvectors of $\mathcal{P}_0$ and $\mathcal{P}_1$, followed by a measurement in the computational basis.

The output of the quantum neuron is defined as
\begin{equation*}
{y} = \left\langle\mathcal{O}\right\rangle = 1-2\left|\bra{1}{U_{\mathcal{O}}} R_z(\gamma) R_y(\beta) R_z(\alpha) \ket{\psi}\right|^2,
\end{equation*}
which represents the average of the observable $\mathcal{O}$. The basis change generated by $U_{\mathcal{O}}$ converts results obtained by measuring Pauli $Z$ into an arbitrary observable $\mathcal{O}$. The accompanying implementation adopts the equivalent opposite eigenvalue convention, corresponding to $\mathcal{O}\mapsto-\mathcal{O}$; this changes only the sign convention for the output. 

Using the definition of the rotation gates, we obtain
\begin{align}\label{eq-yx}
{y} = \, &\cos\beta \cos\theta \cos\omega -{\sin\beta \sin\theta \cos\omega \cos(\alpha+\phi ) }-{\sin\beta \cos\theta \sin\omega \cos(\gamma - \varphi ) }+\nonumber\\
 &\, { \sin\theta\sin\omega \sin(\alpha+\phi ) \sin(\gamma - \varphi ) }-{\cos\beta  \sin \theta\sin\omega  \cos(\alpha+\phi )\cos(\gamma - \varphi ) },
\end{align}
where the angles are considered functions of $x$ with parameters representing weights. With this expression, we can not only obtain good estimates for $\alpha$, $\beta$, and $\gamma$ through the learning process, but also fine-tune the input state and the observable based on the dataset.

All parameters of the circuit of Fig.~\ref{fig-percepton} are present in the expression of the average $\left\langle\mathcal{O}\right\rangle$, represented by ${y}$ in Eq.~\eqref{eq-yx}, including the parameters $\theta$, $\phi$, $\omega$, and $\varphi$ that define the input state $\ket{\psi}$ and the basis rotation. We can set $\phi = \varphi = 0$, without loss of generality, indicating that phases are unnecessary in both the initial state and the projection operators.

\begin{figure}[!h]\label{fig-reduced-neuron}
\centering 
\includegraphics[scale=0.27]{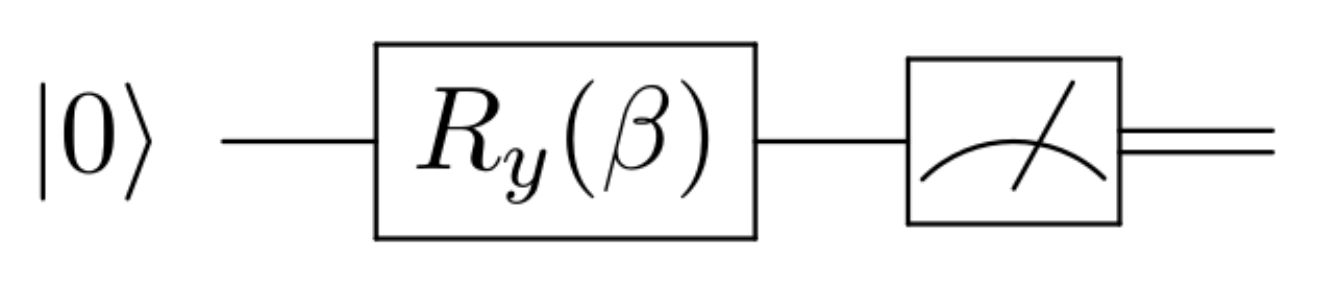}
\caption{Illustration of the reduced quantum neuron with a fixed input state and measurement in the computational basis.}\label{fig-percepton_simplified} 
\end{figure}

Taking $\omega=0$ in Eq.~\eqref{eq-yx}, which corresponds to a measurement in the computational basis, we obtain
\begin{equation}\label{eq-yx2}
{y} = { \cos\beta \cos\theta}-{\cos\alpha \sin\beta  \sin\theta  }.
\end{equation}
In this case, the observable is fixed and cannot be trained. This expression is the most general if we decide to measure in the computational basis. If we further simplify and use $\ket{0}$ as the input to the circuit by setting $\theta = 0$, the outcome of the neuron, which is the expectation value for a measurement in the computational basis, is 
\begin{equation}\label{eq-yx3}    
{y} =1-2\left| \bra{1}R_y(\beta)\ket{0}\right|^2=\cos{\beta}.
\end{equation}
This indicates that, in this case, only the $R_y$ gate, parameterized by a single trainable angle $\beta$, is required. Fig.~\ref{fig-percepton_simplified} shows the reduced version of the quantum neuron, measured in the computational basis.

When measured in the computational basis (\(\omega=0\)), \(y\) simplifies and no longer depends on \(\gamma\), as seen in Eq.~\eqref{eq-yx2}. Furthermore, if the input state is also fixed (\(\theta=0\)), the outcome depends solely on a single real parameter, \(\beta\), as shown in Eq.~\eqref{eq-yx3}. We apply this reduced neuron model to various examples, such as MNIST classification, demonstrating its effectiveness in these cases.

In the next section, we describe a single-qubit quantum neural network that is composed of neurons as depicted in Fig.~\ref{fig-percepton} and a reduced version that is composed of neurons as depicted in Fig.~\ref{fig-percepton_simplified}.

\section{Single-Qubit Quantum Neural Network}\label{sec-QNN}

\begin{figure}[!h]\label{fig-qnn}
\centering 
\includegraphics[scale=0.22]{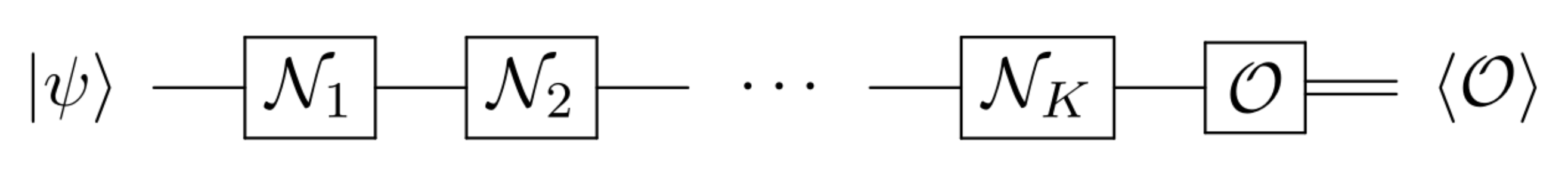}
\caption{Representation of the SQQNN with an arbitrary input and measurement. The output is the average value obtained from measuring the observable $\mathcal{O}$. Each neuron $\mathcal{N}_k$ has three trainable parameters, while the input state $\ket{\psi}$ and the observable $\mathcal{O}$ each have a single trainable parameter, since we have set $\phi=\varphi=0$.} 
\end{figure}

We describe a Single-Qubit Quantum Neural Network (SQQNN) model that uses the interconnection of single-qubit neurons in a streamlined manner. Consider a multi-neuron QNN composed of $K$ neurons ($\mathcal{N}_1$, $\mathcal{N}_2$, $\cdots$, $\mathcal{N}_K$), each embodying a single-qubit architecture as depicted in Fig.~\ref{fig-qnn}. In this configuration, each neuron $\mathcal{N}_k$ is defined by the sequence $\mathcal{N}_k = R_z(\gamma_k) R_y(\beta_k) R_z(\alpha_k)$, for $k = 1, \cdots, K$, where $\alpha_k$, $\beta_k$, and $\gamma_k$ are trainable parameters specific to each neuron. The input state $\ket{\psi}$ and the observable $\mathcal{O}$ each have a single trainable parameter that is not associated with individual neurons. The SQQNN is applicable to both regression and classification tasks.

The SQQNN structure is characterized by the product of the matrices representing each of the $K$ sequentially connected neurons. Since each neuron facilitates an arbitrary 1-qubit rotation, the cumulative effect of combining $K$ neurons is equivalent to a single, effective quantum neuron, as follows:
\[
\mathcal{N}_\text{effective} = \mathcal{N}_K \cdots \mathcal{N}_2 \mathcal{N}_1.
\]
By employing multiple neurons, the model gains flexibility,
increasing the model's expressive power and enhancing its ability to capture complex patterns in the data.

In this design, the angles $\alpha_k$, $\beta_k$, and $\gamma_k$ are determined by arbitrary functions applied to the input data. By composing these functions across multiple neurons, we significantly expand the SQQNN's degrees of freedom and enhance its capacity to adapt to the data during training. Thus, for any function defining the angles in a single-neuron SQQNN, there exists an inverse process and a corresponding set of functions for defining the angles in a multi-neuron SQQNN. This method provides a structured approach to enhance the flexibility of single-neuron SQQNNs, allowing for greater adaptability and improved training outcomes.

The circuit of the network is parameterized by the set of angles $\{\alpha_k, \beta_k, \gamma_k, \theta, \omega\}$. Let $\Omega_k$ denote one of the model's parameters associated with $\mathcal{N}_k$. To implement the learning algorithm, we consider $\Omega_k$ as a function from $\mathbb{R}^p$ to $\mathbb{R}$, representing a trainable weighted function typically derived from dataset inputs. To work with the effective neuron, we extend the function used by a single-neuron model to an effective function $\Omega(x)$, defined as
\begin{equation} 
\label{eq-general_function} 
\Omega(x_i) = c_0 + \sum_{k=1}^K \sum_{j=1}^p c_{kj} x_{ij}^k, 
\end{equation} 
where $c_{kj}$ represents the unknown coefficients to be determined during training; $K$ indicates the number of power terms in the formal construction; and $x_{ij}$ is the $j$-th component of dataset instance $x_i$, which has $p$ dimensions. The argument of $\Omega(x)$ can be an arbitrary vector $x$ in $\mathbb{R}^p$. Although Eq.~\eqref{eq-general_function} is a specific choice, alternative formulations are possible. This additive polynomial ansatz provides progressively richer dependence on individual features; cross-feature monomials can be added when needed. It is derived from the reduced network, where the rotation parameters sum to give Eq.~\eqref{eq-general_function}.

\begin{figure}[!h]
\includegraphics[scale=0.22]{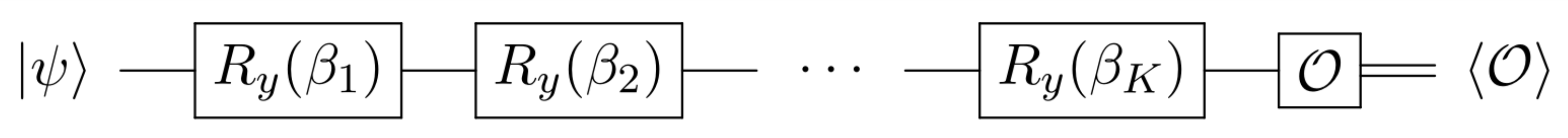}
\caption{Representation of the reduced SQQNN. Each neuron $\mathcal{N}_k$ is the rotation operator $R_y(\beta_k)$, which has only one trainable parameter.} \label{fig-reduced}
\end{figure}

We propose a reduced SQQNN defined by using the reduced neuron with the activation function~\eqref{eq-yx3} as illustrated in Fig.~\ref{fig-reduced-neuron}. In this setup, the network's circuit comprises a sequence of $R_y(\beta_k)$ rotations, each parameterized by angle $\Omega_k=\beta_k$, defined as
\begin{equation*}
\beta_k(x_i) = c_0\delta_{k1} + \sum_{j=1}^p c_{kj} x_{ij}^k,
\end{equation*}
where $\delta$ is the Kronecker delta. The network is depicted in Fig.~\ref{fig-reduced}. The collective effect of all neurons is equivalent to a single $R_y(\beta)$ rotation, where all angles are summed as $\beta=\sum_{k=1}^K\beta_k$, leading to Eq.~\eqref{eq-general_function}. Consequently, the SQQNN model simplifies to the circuit proposed in Fig.~\ref{fig-percepton_simplified}.

Our proposal differs from Ref.~\cite{Salinas2020} in at least two key aspects. First, in Ref.~\cite{Salinas2020}, the input data are directly used as the arguments of three-parameter unitary gates, requiring no classical pre-processing when the data have exactly three dimensions. In contrast, our approach can handle input data with any number of dimensions, but it requires classical pre-processing to compute the final values of the angles used as arguments for the rotation gates. Second, the algebraic operation intrinsic to the model in Ref.~\cite{Salinas2020} is the Hadamard product between the weights and the input data, whereas in our approach, a single neuron computes the dot product between the weights and the input data. For architectures involving multiple neurons, the dot product extends to the weights and the powers of the input data entries.

We consistently use the functional form in Eq.~\eqref{eq-general_function} in the applications below. In the mathematical construction, $K$ denotes the maximum polynomial degree. To distinguish the indexing convention used by the released software, we write its model-order setting as $K_{\mathrm{code}}$: preprocessing constructs powers $x_{ij}^k$ for $k=1,\ldots,K_{\mathrm{code}}+1$. Therefore, $K_{\mathrm{code}}$ in the experimental tables denotes a software model-order setting rather than the number of non-redundant quantum gates. As shown above, the reduced sequence can still be compiled into a single $R_y$ rotation.

\section{Training Methods}\label{sec-LM}

In machine learning, training refers to the process through which a model improves its performance on a specific task by identifying patterns within a dataset. This improvement is typically achieved by optimizing a set of weights,  $(c_0,...,c_p)\in \mathbb{R}^{p+1}$, with the goal of minimizing the loss function. This function quantifies the discrepancy between the model's predictions and the actual outcomes. Throughout training, the model iteratively adjusts its parameters based on feedback from the loss function, gradually reducing errors and enhancing predictive accuracy. This process is generally carried out through gradient descent, which incrementally refines the model's parameters by following the gradient of the loss function, or via other optimization algorithms described later. The ultimate goal of learning is to enable the model to generalize well, not only on the training data but also on new, unseen data, by capturing essential patterns and relationships within the dataset.

For regression tasks, we use the Mean Square Error (MSE) loss function
\begin{equation*}
\text{MSE}(\Omega,\mathbb{D}) = \frac{1}{n}\sum_{i=1}^n \left(\hat{y}_i-y_i\right)^2,
\end{equation*} 
and for classification tasks, we use the hinge loss function
\begin{equation*}
\text{Hinge}(\Omega,\mathbb{D}) = \frac{1}{n}\sum_{i=1}^n \max\{0,1-\hat{y}_iy_i\},
\end{equation*} 
where $\hat{y}_i$ is the model's estimated output when the input is the dataset element $x_i$. The estimated output $\hat{y}_i$ is calculated using Eq.~\eqref{eq-yx}. With $\phi$ and $\varphi$ set to 0, the model's output is defined as follows
\begin{align}\label{eq-yx2a}
\hat{y} =&\cos\beta \cos\theta \cos\omega -\cos\alpha\sin\beta \sin\theta \cos\omega  - \sin\beta \cos\gamma \cos\theta \sin\omega  \, +\nonumber\\
 &\,   \sin\alpha \sin\gamma\sin\theta\sin\omega  - \cos\alpha\cos\beta \cos\gamma \sin \theta\sin\omega  .
\end{align}
The angles $\alpha$, $\beta$, $\gamma$, $\theta$, and $\omega$ are estimated based on the weights and input data. The loss function can be minimized with respect to the weights using standard methods such as the gradient descent method.

\subsection{Gradient Descent}

The primary concept of gradient descent is to iteratively adjust the model parameters in the opposite direction of the gradient of the loss function. Initially, the model parameters are set randomly. In each iteration, the gradient of the loss function with respect to the parameters is computed. The parameters are then updated by a factor controlled by the learning rate ($\eta$), as shown in the following equation:
\begin{equation*}
\Omega_{t+1}=\Omega_t-\eta\nabla_{\Omega}\text{Loss}\left(\Omega_t,\mathbb{D} \right).
\end{equation*}
This process continues until the loss function reaches a sufficiently low value or until the gradient approaches zero, indicating that a minimum has been reached. The advantages of using gradient descent include simplicity of the model, efficiency, and adaptability to high-dimensional data. However, it can become trapped in local minima. It is important to choose a learning rate that ensures effective convergence.

To apply the gradient descent method for the single-qubit case, it is necessary to calculate the gradient of the loss function with respect to the angles $\alpha$, $\beta$, $\gamma$, $\theta$, and $\omega$. Eventually, we analytically compute the partial derivatives of 
    $\hat{y}$ (Eq.~\eqref{eq-yx2a})  with respect to these angles, which simplifies the training process. The expressions for these derivatives are
\begin{align*}
{\partial \hat{y}}/{\partial \alpha} \,=\, & 
 \sin\alpha\sin\beta \sin\theta \cos\omega  +  \cos\alpha \sin\gamma\sin\theta\sin\omega  + \sin\alpha\cos\beta \cos\gamma \sin \theta\sin\omega;
\\
{\partial \hat{y}}/{\partial \beta} \,=\, & 
-\sin\beta \cos\theta \cos\omega -\cos\alpha\cos\beta \sin\theta \cos\omega  - \cos\beta \cos\gamma \cos\theta \sin\omega \, + \\ & \cos\alpha\sin\beta \cos\gamma \sin \theta\sin\omega;
\\
{\partial \hat{y}}/{\partial \gamma} \,=\, & 
 \sin\beta \sin\gamma \cos\theta \sin\omega  +  \sin\alpha \cos\gamma\sin\theta\sin\omega  + \cos\alpha\cos\beta \sin\gamma \sin \theta\sin\omega;
\\
{\partial \hat{y}}/{\partial \theta} \,=\, & 
-\cos\beta \sin\theta \cos\omega -\cos\alpha\sin\beta \cos\theta \cos\omega  + \sin\beta \cos\gamma \sin\theta \sin\omega \, + \\ & \sin\alpha \sin\gamma\cos\theta\sin\omega  - \cos\alpha\cos\beta \cos\gamma \cos \theta\sin\omega;
\\
{\partial \hat{y}}/{\partial \omega} \,=\, & 
-\cos\beta \cos\theta \sin\omega +\cos\alpha\sin\beta \sin\theta \sin\omega  - \sin\beta \cos\gamma \cos\theta \cos\omega \, + \\ & \sin\alpha \sin\gamma\sin\theta\cos\omega  - \cos\alpha\cos\beta \cos\gamma \sin \theta\cos\omega.
\end{align*}

\subsection{Polynomial-based Linear Least Squares}

In this subsection, we introduce a training algorithm specifically designed for binary classification tasks using reduced SQQNNs, based on successive powers of the entries of the dataset elements. The function mapping data to the angle $\beta$ is defined as
\begin{equation*} 
\beta(x_i) = \arccos\left[ \tanh\left(c_0 + \sum_{k=1}^K \sum_{j=1}^p c_{kj} x_{ij}^k \right)\right].
\end{equation*} 
Substituting this function into Eq.~\eqref{eq-yx3} yields
\begin{equation*} 
\text{arctanh}(y'_i) = c_0 + \sum_{k=1}^K \sum_{j=1}^p c_{kj} x_{ij}^k,
\end{equation*} 
where $y'_i = -1 + \epsilon$ if $y_i = -1$, $y'_i = 1 - \epsilon$ if $y_i = 1$, and $y'_i=y_i$ otherwise. The small parameter $\epsilon > 0$ is introduced because the $\tanh$ function never reaches -1 or 1, necessitating the adjustment of $y_i$ by this parameter. Typically, in applications, we set $\epsilon = 10^{-16}$.

The coefficients $c_0$ and $c_{kj}$ are determined using the least squares linear regression formula
\begin{equation} 
\label{eq-linear_regression}
S = (X^T X)^{+} X^T Y,
\end{equation} 
where \( S \) is the vector of coefficients
\[
{S} = \begin{bmatrix} c_0 & c_{11} & \cdots & c_{1p} & \cdots & c_{K1} & \cdots & c_{Kp} \end{bmatrix}^T,
\]
\( Y \) is the label vector modified by the \(\text{arctanh}\) function
\[
{Y} = \begin{bmatrix} \text{arctanh}(y'_1) & \text{arctanh}(y'_2) & \cdots & \text{arctanh}(y'_n) \end{bmatrix}^T,
\]
and \( X \) is a matrix constructed from the dataset inputs
\begin{displaymath}
{X}=\left[\begin{array}{cccccccccccc}
1 & {x}_{11} & \cdots & {x}_{1p} & \cdots & {x}_{11}^k & \cdots & {x}_{1p}^k & \cdots & {x}_{11}^K & \cdots & {x}_{1p}^K \\
\vdots  & \vdots  & \vdots  & \vdots & \vdots  & \vdots  & \vdots  & \vdots  & \vdots  & \vdots  & \vdots  & \vdots  \\
1 & {x}_{i1} & \cdots & {x}_{ip} & \cdots & {x}_{i1}^k & \cdots & {x}_{ip}^k & \cdots & {x}_{i1}^K & \cdots & {x}_{ip}^K\\
\vdots  & \vdots  & \vdots  & \vdots & \vdots  & \vdots  & \vdots  & \vdots  & \vdots  & \vdots  & \vdots  & \vdots  \\
1 & {x}_{n1} & \cdots & {x}_{np} & \cdots & {x}_{n1}^k & \cdots & {x}_{np}^k & \cdots & {x}_{n1}^K & \cdots & {x}_{np}^K\\
\end{array}\right].
\end{displaymath}
\( X \) includes not only the raw dataset inputs but also the successive powers of their entries up to degree \( K \), enabling the model to capture higher-order relationships in the data.

In this formulation, \( X^T \) represents the transpose of the matrix \( X \), and \( (X^T X)^{+} \) denotes the Moore--Penrose pseudo-inverse of the matrix \( X^T X \). This approach allows for the efficient determination of network parameters in a single step, significantly accelerating the training process. The procedure returns a global minimizer of the transformed least-squares objective.

The polynomial-based linear least squares method has two key advantages for training reduced SQQNNs. First, it admits a closed-form solution (Eq.~\eqref{eq-linear_regression}), enabling fast and deterministic parameter estimation in a single step, without the need for iterative updates. Second, because the loss function is quadratic and the system is solved analytically, the closed-form solution is a global minimizer of the transformed quadratic least-squares objective~\cite{Bishop2006}, avoiding local minima that may arise with gradient-based training. By including polynomial terms up to degree $K$, the model captures nonlinear relationships while maintaining computational simplicity.

\section{Regression Evaluation}\label{sec-Regression}

In this section, we evaluate the effectiveness of SQQNN in performing regression tasks. Regression is a critical application for quantum neural networks as it requires accurately modeling continuous relationships and predicting real-valued outputs,  which were initially rescaled to the range [-1, 1] and subsequently recalibrated to their original range. We apply SQQNN to three types of regression tasks: logic gates, modeling continuous functions, and analyzing real-world datasets. By assessing its performance in these cases, we aim to demonstrate SQQNN's precision and adaptability in diverse regression scenarios, using the gradient descent method. For the real-world datasets, we employed 10-fold cross-validation to ensure robust evaluation.

\subsection{Logic Gate Evaluation}\label{sec-logic_gates}

The results of training the SQQNN to evaluate logical gates are shown in Table~\ref{tab-logic_gates}. Here, we present the training error and the number of gradient descent epochs required for each gate, including AND, OR, XOR, NAND, NOR, and XNOR. The SQQNN used a single effective neuron with software model-order setting $K_{\mathrm{code}}=1$.

\begin{table}[!ht]
\caption{Experimental results of training the SQQNN on logical gates using gradient descent.}
\centering 
{\small
\renewcommand{\arraystretch}{1.0}
\begin{tabular}{|c|c|c|c|c|c|}
\hline
\textbf{Gate} & \textbf{MSE} & \textbf{Epochs} & \textbf{Gate} & \textbf{MSE} & \textbf{Epochs} \\
\hline
AND   & \(9 \cdot 10^{-4}\) & 38 & NAND & \(2 \cdot 10^{-3}\) & 61 \\
\hline
OR    & \(2 \cdot 10^{-4}\) & 64 & NOR  & \(2 \cdot 10^{-3}\) & 75 \\
\hline
XOR   & \(1 \cdot 10^{-3}\) & 79 & XNOR & \(6 \cdot 10^{-4}\) & 47 \\
\hline
\end{tabular}
}
\label{tab-logic_gates}
\end{table}

Table~\ref{tab-logic_gates} shows that SQQNN achieved low training errors across all gates, indicating effective error minimization. The number of gradient descent iterations also remained relatively low across all logical gates, including more complex, nonlinear ones such as XOR and XNOR. These results highlight the SQQNN's ability to learn efficiently across various logical functions, demonstrating its flexibility in capturing both linear and non-linear behavior in binary logic operations.



Ref.~\cite{Nguyen2020} analyzed logic gates using two-qubit QNNs and reported that the QNN achieved less than $1\%$ error within a single epoch. Our experiments indicate that the SQQNN requires at least 5 epochs to reach an error rate below $1\%$. In contrast, the classical neural networks analyzed in Ref.~\cite{Nguyen2020} required substantially more training epochs for the same task. For the logic-gate perceptron experiments, real-valued classical networks required several thousand epochs (typically about $6\times10^3$--$1.1\times10^4$) to reach an RMS error of $1\%$, while complex-valued networks required roughly $10^2$ epochs. These epoch counts are much larger than those required by the quantum models, including the SQQNN results summarized in Table~\ref{tab-logic_gates}, which also achieved small errors. This comparison suggests that, for certain regression tasks, two-qubit QNNs may train faster than single-qubit QNNs.

\subsection{Sinc function}

The sinc function, defined as 
\begin{equation*} 
\text{sinc}(x) = \frac{\sin(x)}{x}, 
\end{equation*} 
is a standard benchmark for regression tasks due to its oscillatory, non-linear nature and decaying amplitude, which present challenges for models to accurately capture without overfitting. For this experiment, we generated data based on the sinc function and trained SQQNNs with a single neuron to predict previously unseen data points. The dataset consisted of 800 data points for training, 100 for validation, and 100 for testing. 

\begin{figure}[!h]
\centering 
\includegraphics[scale=0.38]{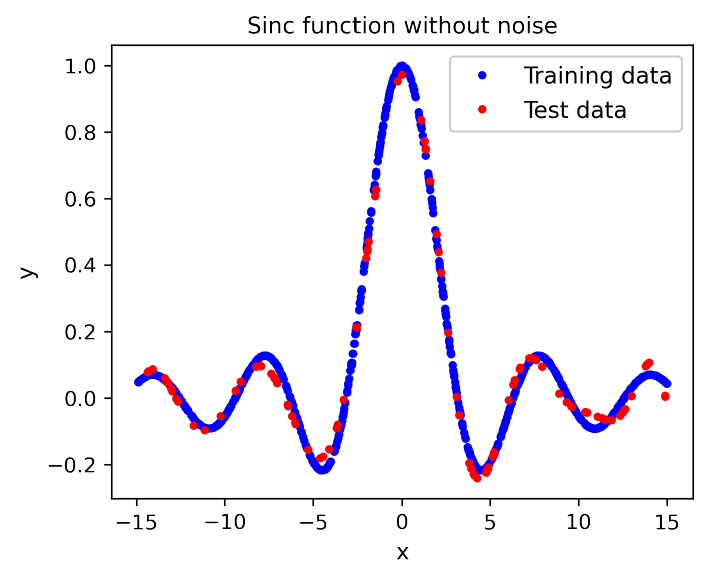}
\includegraphics[scale=0.38]{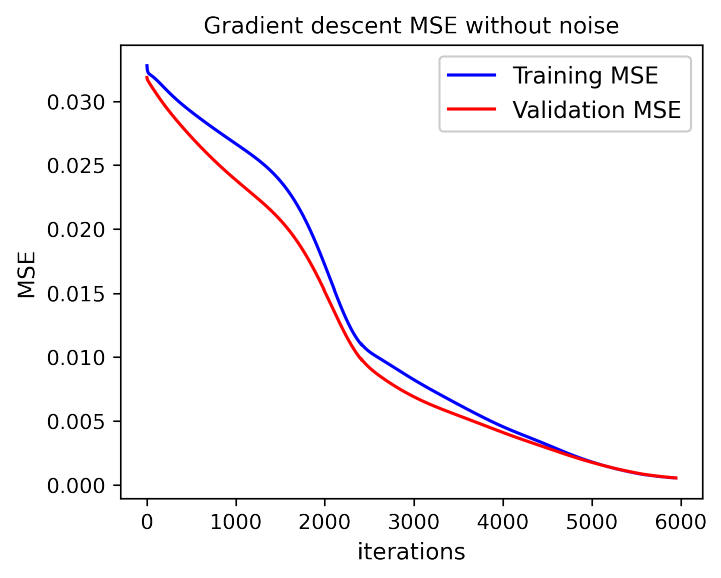}\\
\includegraphics[scale=0.38]{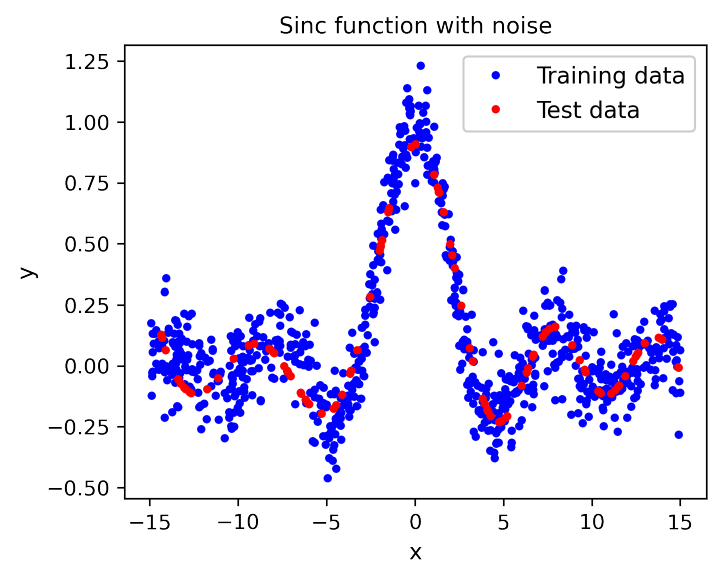}
\includegraphics[scale=0.38]{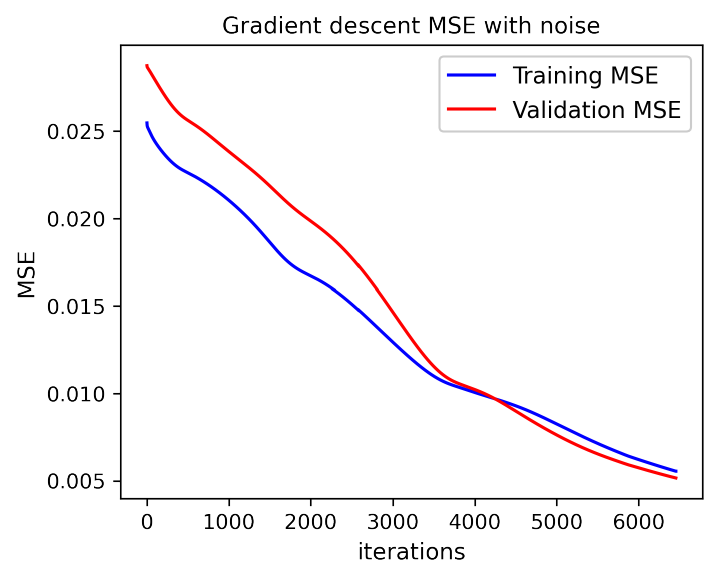}
\caption{Performance of the SQQNN on a sinc function regression task using a single neuron. The first panel shows the combined training and test data without noise, while the second panel presents the corresponding error as a function of the number of iterations. The third and fourth panels show the same results after adding white noise to the data.}
\label{fig-sinc} 
\end{figure}

Figure~\ref{fig-sinc} illustrates the performance of the SQQNN on this task under two scenarios: training with noiseless data and training with data affected by white noise with standard deviation of 0.01. In the top-left panel, the training and test datasets for the noiseless case are shown, clearly aligning with the sinc function, while the corresponding mean squared error, plotted as a function of iterations in the top-right panel, converges to approximately $10^{-3}$ after $7 \cdot 10^4$ iterations. Similarly, the bottom-left panel shows the training and test datasets when noise is added to the data, with the SQQNN successfully approximating the sinc function despite the noise. The bottom-right panel demonstrates that the error decreases at a comparable rate, even with noisy training data. These results highlight the SQQNN's robustness and ability to handle both clean and noisy regression tasks effectively.

\subsection{Combined Cycle Power Plant Dataset}

The Combined Cycle Power Plant (CCPP) dataset, sourced from the UCI Machine Learning Repository\footnote{\url{https://archive.ics.uci.edu/dataset/294/combined+cycle+power+plant}}, provides 9,568 samples of data collected from a gas turbine under full load conditions. It features four input variables: ambient temperature, ambient pressure, relative humidity, and exhaust vacuum; and one target variable, the net electrical energy output of the plant, measured in megawatts. The dataset is widely used for regression tasks due to its smooth, non-linear relationships and strong correlations between features and the target variable. With its clean and comprehensive structure, the CCPP dataset serves as a valuable benchmark for modeling and predicting energy production under varying environmental and operational conditions.

\begin{table}[!h]
\caption{Performance metrics (MSE for training, validation, and test) for the combined cycle power plant dataset as the software model-order setting $K_{\mathrm{code}}$ increases.}
\centering 
{\small
\renewcommand{\arraystretch}{1.0}
\begin{tabular}{|c|c|c|c|}
\hline
 \textbf{Model order $K_{\mathrm{code}}$} & \textbf{Training MSE} & \textbf{Validation MSE} & \textbf{Test MSE} \\
\hline
 1 & 0.0059 $\pm$ 0.0008 & 0.0059 $\pm$ 0.0008 & 0.0059 $\pm$ 0.0007 \\
2 & 0.0045 $\pm$ 0.0005 & 0.0048 $\pm$ 0.0006 & 0.0045 $\pm$ 0.0005 \\
 3 & 0.0040 $\pm$ 0.0002 & 0.0043 $\pm$ 0.0003 & 0.0040 $\pm$ 0.0004 \\
4 & 0.0037 $\pm$ 0.0002 & 0.0042 $\pm$ 0.0003 & 0.0037 $\pm$ 0.0003 \\
5 & 0.0038 $\pm$ 0.0004 & 0.0043 $\pm$ 0.0004 & 0.0039 $\pm$ 0.0004 \\
6 & 0.0035 $\pm$ 0.0002 & 0.0040 $\pm$ 0.0003 & 0.0035 $\pm$ 0.0002 \\
\hline
\end{tabular}
}\label{tab-comb-cycle-power}
\end{table}

Table~\ref{tab-comb-cycle-power} presents the performance of the model on the Combined Cycle Power Plant dataset as the software model-order setting $K_{\mathrm{code}}$ increases. The results show an overall improvement in MSE up to $K_{\mathrm{code}}=6$, with the test MSE decreasing from $0.0059 \pm 0.0007$ for $K_{\mathrm{code}}=1$ to $0.0035 \pm 0.0002$ for $K_{\mathrm{code}}=6$. This indicates that increasing the tested model order can improve the model's ability to capture non-linear relationships in this dataset. The small standard deviations suggest stable performance across the evaluated subsets.

\subsection{Communities and Crime Dataset}

The Communities and Crime dataset, sourced from the UCI Machine Learning Repository\footnote{\url{https://archive.ics.uci.edu/dataset/183/communities+and+crime}}, provides comprehensive data on crime rates across 1,994 U.S. communities, alongside key socio-economic and demographic factors. The dataset focuses on various types of crimes, including violent offenses such as murder, rape, robbery, and assault, as well as property crimes like burglary, theft, and motor vehicle theft. Its target variable is the violent crime rate per capita, making it particularly suitable for regression analysis. In addition to crime statistics, the dataset includes socio-economic variables such as population size, income levels, education attainment, poverty rates, and the proportion of minority populations. With its combination of detailed crime metrics and contextual information, this dataset enables an in-depth exploration of the relationships between community characteristics and crime rates, as well as regional variations in crime patterns. This dataset was used in many papers~\cite{REDMOND2002660,McClendon2015,Kolomoytseva2021}.

\begin{table}[!h]
\caption{Performance metrics for the communities and crime dataset as the software model-order setting $K_{\mathrm{code}}$ increases.}
\centering 
{\small
\renewcommand{\arraystretch}{1.0}
\begin{tabular}{|c|c|c|c|}
\hline
 \textbf{Model order $K_{\mathrm{code}}$} & \textbf{Training MSE} & \textbf{Validation MSE} & \textbf{Test MSE} \\
\hline
 1 & 0.0426 $\pm$ 0.0111 & 0.0405 $\pm$ 0.0110 & 0.0434 $\pm$ 0.0133 \\
2 & 0.0522 $\pm$ 0.0099 & 0.0561 $\pm$ 0.0151 & 0.0539 $\pm$ 0.0154 \\
 3 & 0.0604 $\pm$ 0.0239 & 0.0613 $\pm$ 0.0260 & 0.0600 $\pm$ 0.0235 \\
4 & 0.0733 $\pm$ 0.0203 & 0.0842 $\pm$ 0.0298 & 0.0804 $\pm$ 0.0235 \\
5 & 0.0853 $\pm$ 0.0227 & 0.0833 $\pm$ 0.0274 & 0.0853 $\pm$ 0.0225 \\
6 & 0.1168 $\pm$ 0.0242 & 0.1234 $\pm$ 0.0202 & 0.1195 $\pm$ 0.0252 \\
\hline
\end{tabular}
}
\label{tab-comm-and-crimes}
\end{table}

Table~\ref{tab-comm-and-crimes} summarizes the performance of SQQNN on the Communities and Crime dataset for varying software model-order settings $K_{\mathrm{code}}$. The training, validation, and test MSE values increase as $K_{\mathrm{code}}$ grows, with the lowest MSE observed for $K_{\mathrm{code}}=1$ (training MSE: $0.0426 \pm 0.0111$, test MSE: $0.0434 \pm 0.0133$). Because the training error also increases, this trend does not by itself establish overfitting; rather, the higher tested orders did not improve performance for this dataset. The results therefore favor the smaller model-order setting.

\section{Classification Evaluation}\label{sec-Class}

In this section, we present the results of the SQQNN architecture, evaluated on both synthetic and real datasets to demonstrate its effectiveness and versatility in various binary classification tasks. The training method employed is polynomial-based linear least squares. For the analysis of real datasets, we employed accuracy, precision, sensitivity, and F1 score to evaluate classification performance. The output classes used 1 and -1 to represent each category. Additionally, 10-fold cross-validation was applied to the real datasets.

\subsection{Two Moons Dataset}

The Two Moons dataset is a synthetic dataset commonly used for testing classification models. It consists of two interlocking half-moon shapes that are challenging to separate linearly, making it an ideal benchmark for assessing the capabilities of nonlinear classifiers. For this experiment, the dataset was generated with a noise parameter set to 0.07; the training set comprised 1,000 data points, and the test set comprised 100 samples.

\begin{table}[!h]
\caption{Performance metrics for the Two Moons dataset as the software model-order setting $K_{\mathrm{code}}$ increases.}
\centering 
{\small
\renewcommand{\arraystretch}{1.0}
\begin{tabular}{|c|c|c|c|c|c|c|}
\hline
 \textbf{Model order $K_{\mathrm{code}}$}  & \textbf{Accuracy} & \textbf{Precision} & \textbf{Sensitivity} & \textbf{Specificity} & \textbf{F1 Score} \\
\hline
1  & 0.87 & 0.8776 & 0.86 & 0.88 & 0.8687 \\
 2 &  0.99 & 0.9804 & 1.0 & 0.98 & 0.9901 \\
3  & 0.99 & 0.9804 & 1.0 & 0.98 & 0.9901 \\
 4  & 1.0 & 1.0 & 1.0 & 1.0 & 1.0 \\
5 & 1.0 & 1.0 & 1.0 & 1.0 & 1.0 \\
 6  & 1.0 & 1.0 & 1.0 & 1.0 & 1.0 \\
\hline
\end{tabular}
}
\label{tab-moon-class-res}
\end{table}

\begin{figure}[!h]
\centering 
\includegraphics[scale=0.45]{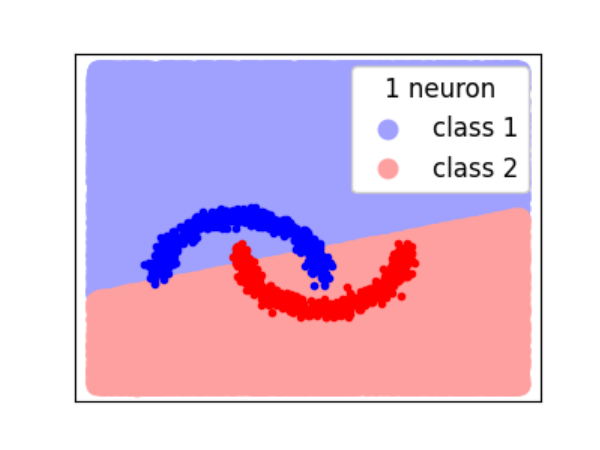}\includegraphics[scale=0.45]{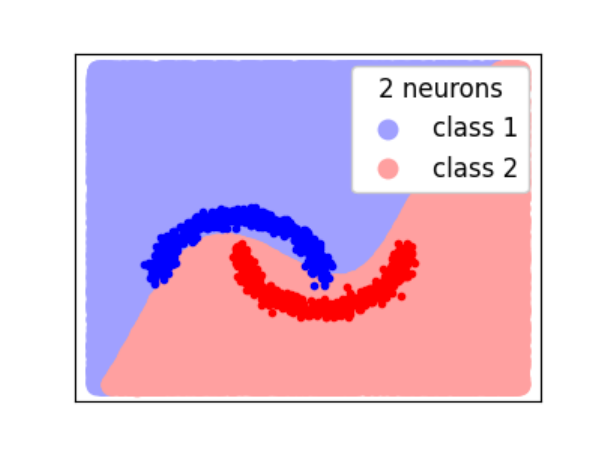}\includegraphics[scale=0.45]{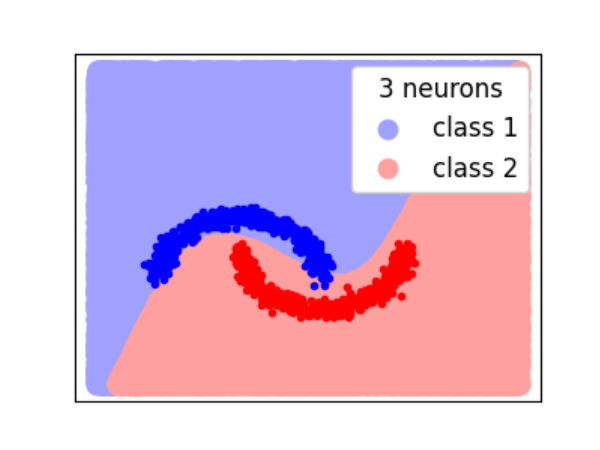}\\
\includegraphics[scale=0.45]{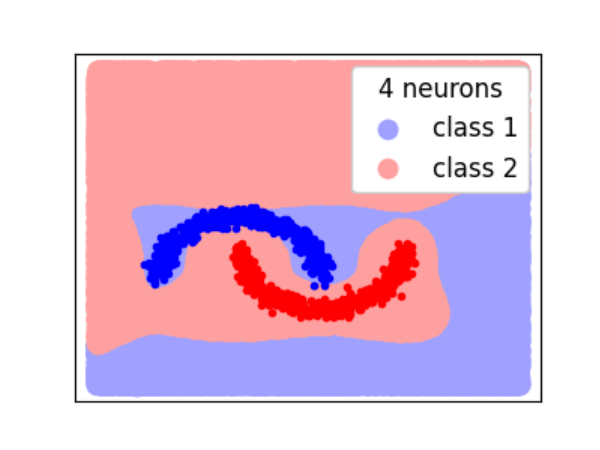}\includegraphics[scale=0.45]{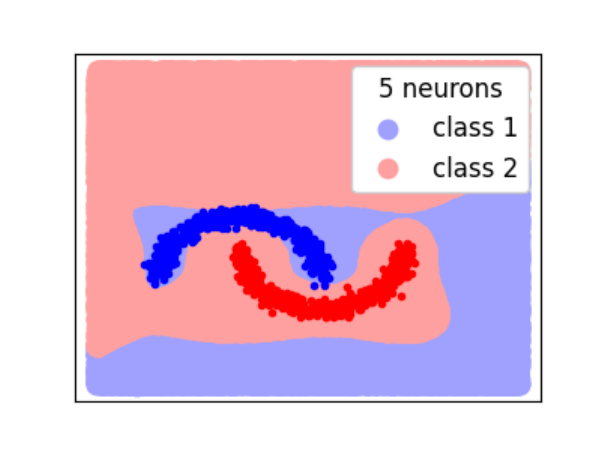}\includegraphics[scale=0.45]{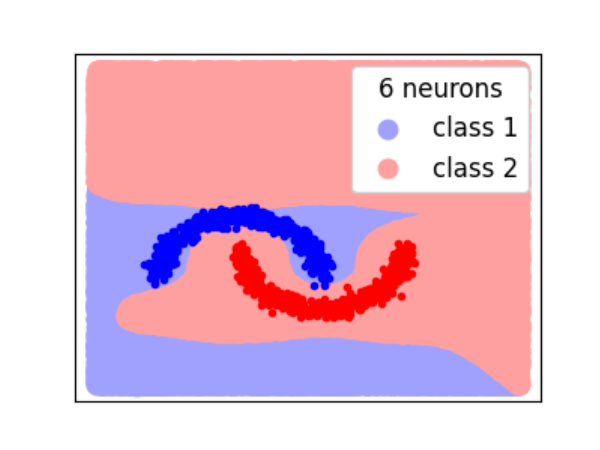}
\caption{Decision boundaries of the SQQNN classifier for software model-order settings $K_{\mathrm{code}}=1$ to 6 on the two-dimensional, nonlinearly separable Two Moons dataset.}\label{fig-moons} \end{figure}

The results in Table~\ref{tab-moon-class-res} show that performance improves from an accuracy of 0.87 at $K_{\mathrm{code}}=1$ to 0.99 at $K_{\mathrm{code}}=2$ and 3, reaching 1.0 for the tested settings $K_{\mathrm{code}}=4$, 5, and 6. Fig.~\ref{fig-moons} illustrates this progression through increasingly flexible decision boundaries on the nonlinearly separable Two Moons dataset.

\subsection{Wisconsin Breast Cancer Dataset}

The Wisconsin Breast Cancer Dataset (WBCD), sourced from the UCI Machine Learning Repository\footnote{\url{https://archive.ics.uci.edu/dataset/15/breast+cancer+wisconsin+original}}, is a widely recognized benchmark in medical diagnostics and machine learning research. This dataset comprises 569 samples, each characterized by 30 numeric features derived from fine needle aspiration (FNA) tests of breast tissue. The features capture various morphological properties of cell nuclei, including radius, texture, perimeter, area, and smoothness, among others. These descriptors enable the differentiation between benign and malignant tumors, facilitating accurate diagnostic predictions. The WBCD serves as a critical testbed for evaluating the performance of classification algorithms, offering a well-balanced distribution of classes and a real-world application of machine learning in healthcare.

The results presented in Table~\ref{tab-wisconsin} were produced using software model-order settings $K_{\mathrm{code}}=1$ to 10. By progressively increasing $K_{\mathrm{code}}$, we evaluated the relationship between model complexity and classification performance on the WBCD dataset.

\begin{table}[!h]
\caption{Performance metrics for the WBCD dataset as the software model-order setting $K_{\mathrm{code}}$ increases.}
\centering 
{\scriptsize
\renewcommand{\arraystretch}{0.9}
\begin{tabular}{|c|c|c|c|c|}
\hline
 \textbf{Model order $K_{\mathrm{code}}$} & \textbf{Accuracy (\(\pm\))} & \textbf{Precision (\(\pm\))} & \textbf{Sensitivity (\(\pm\))} & \textbf{F1 (\(\pm\))} \\
\hline
 1  & 0.951 \(\pm\) 0.025 & 0.994 \(\pm\) 0.017 & 0.878 \(\pm\) 0.068 & 0.931 \(\pm\) 0.034 \\
 2  & 0.954 \(\pm\) 0.020 & 1.000 \(\pm\) 0.000 & 0.879 \(\pm\) 0.047 & 0.935 \(\pm\) 0.027 \\
 3 & 0.954 \(\pm\) 0.021 & 1.000 \(\pm\) 0.000 & 0.880 \(\pm\) 0.050 & 0.935 \(\pm\) 0.029 \\
 4  & 0.956 \(\pm\) 0.023 & 0.988 \(\pm\) 0.024 & 0.894 \(\pm\) 0.052 & 0.938 \(\pm\) 0.034 \\
 5 & 0.954 \(\pm\) 0.027 & 0.988 \(\pm\) 0.025 & 0.890 \(\pm\) 0.054 & 0.936 \(\pm\) 0.039 \\
 6  & 0.956 \(\pm\) 0.032 & 0.987 \(\pm\) 0.026 & 0.896 \(\pm\) 0.074 & 0.938 \(\pm\) 0.050 \\
 7 & 0.949 \(\pm\) 0.030 & 0.982 \(\pm\) 0.028 & 0.882 \(\pm\) 0.072 & 0.928 \(\pm\) 0.045 \\
 8  & 0.958 \(\pm\) 0.028 & 0.982 \(\pm\) 0.028 & 0.905 \(\pm\) 0.066 & 0.941 \(\pm\) 0.042 \\
 9 & 0.958 \(\pm\) 0.028 & 0.982 \(\pm\) 0.028 & 0.905 \(\pm\) 0.066 & 0.941 \(\pm\) 0.042 \\
 10 & 0.956 \(\pm\) 0.028 & 0.982 \(\pm\) 0.028 & 0.901 \(\pm\) 0.066 & 0.938 \(\pm\) 0.041 \\
\hline
\end{tabular}
}\label{tab-wisconsin}
\end{table}

Despite fluctuations in the performance measures, Table~\ref{tab-wisconsin} shows a modest overall improvement as the software model-order setting increases. The consistently low standard deviations indicate stable performance across folds. Increasing $K_{\mathrm{code}}$ beyond 8 provides no further improvement in the reported accuracy.

\subsection{MNIST Dataset}

The MNIST dataset, a standard benchmark in the field of machine learning, consists of 70,000 grayscale images of handwritten digits, partitioned into 60,000 training samples and 10,000 test samples. Each image, represented as a $28\times 28$ pixel grid, encapsulates the complexity of real-world digit recognition tasks while maintaining computational simplicity. Despite its relative simplicity, MNIST remains a critical testbed for evaluating novel neural network architectures. In this work, we use MNIST to show the performance and generalization potential of the SQQNN architecture, establishing a baseline for its capabilities in image recognition tasks. 

To control input dimensionality and reduce the number of trainable parameters, we applied the Discrete Cosine Transform (DCT) to each image as a pre-processing step and vectorized the resulting DCT matrix using the Zig-Zag traversal method~\cite{gonzalez2018digital}. This method arranges the coefficients by frequency, placing the most informative low-frequency components first. This ordering facilitates compression, as many high-frequency coefficients tend to be close to zero and can be discarded. For all experiments, we used the first 50 coefficients as input to the network.

 \begin{figure}[!h]
     \centering 
     \includegraphics[width=0.5\linewidth]{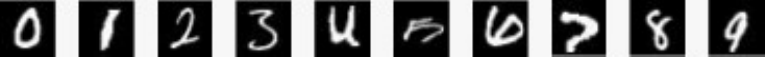}
     \caption{Examples of digits from the MNIST dataset incorrectly predicted.}
     \label{fig-inc-prediction}
 \end{figure}

Fig.~\ref{fig-inc-prediction} depicts sample digits from 0 to 9 from the MNIST dataset that were incorrectly predicted, highlighting the variability in handwriting styles and pixel intensity. 

The results presented in Table~\ref{tab-mnist} were produced using the reduced SQQNN with software model-order setting $K_{\mathrm{code}}=1$, specifically designed to optimize computational efficiency while maintaining high classification accuracy. The reduced model was trained using polynomial-based linear least squares on the MNIST training set. Each training iteration took approximately 8 seconds on a DELL Vostro 3480 notebook with an Intel Core i5-8265U 1.60GHz x8 CPU, 8GB DDR4 RAM, and Ubuntu 22.04.4 LTS.

\begin{table}[!t]
\caption{MNIST performance metrics for all class pairings using the reduced SQQNN with software model-order setting $K_{\mathrm{code}}=1$.}
\centering 
{\scriptsize
\renewcommand{\arraystretch}{0.8}
\begin{tabular}{|c|c|c|c|c|}
\hline
 \textbf{Classes} & \textbf{Accuracy} & \textbf{Precision} & \textbf{Sensitivity} & \textbf{F1} \\
\hline
 0 and 1 & 0.999 $\pm$ 0.001 & 0.999 $\pm$ 0.001 & 0.999 $\pm$ 0.001 & 0.999 $\pm$ 0.001 \\
 0 and 2 & 0.995 $\pm$ 0.002 & 0.996 $\pm$ 0.002 & 0.994 $\pm$ 0.002 & 0.995 $\pm$ 0.002 \\
 0 and 3 & 0.997 $\pm$ 0.002 & 0.997 $\pm$ 0.002 & 0.997 $\pm$ 0.003 & 0.997 $\pm$ 0.002 \\
  0 and 4 & 0.997 $\pm$ 0.001 & 0.996 $\pm$ 0.002 & 0.997 $\pm$ 0.002 & 0.997 $\pm$ 0.001 \\
 0 and 5 & 0.995 $\pm$ 0.002 & 0.995 $\pm$ 0.003 & 0.995 $\pm$ 0.002 & 0.995 $\pm$ 0.002 \\
 0 and 6 & 0.993 $\pm$ 0.002 & 0.993 $\pm$ 0.003 & 0.993 $\pm$ 0.002 & 0.993 $\pm$ 0.002 \\
0 and 7 & 0.997 $\pm$ 0.001 & 0.996 $\pm$ 0.003 & 0.998 $\pm$ 0.002 & 0.997 $\pm$ 0.001 \\
 0 and 8 & 0.995 $\pm$ 0.002 & 0.996 $\pm$ 0.003 & 0.994 $\pm$ 0.003 & 0.995 $\pm$ 0.002 \\
 0 and 9 & 0.993 $\pm$ 0.003 & 0.995 $\pm$ 0.002 & 0.992 $\pm$ 0.004 & 0.993 $\pm$ 0.003 \\
  1 and 2 & 0.992 $\pm$ 0.002 & 0.991 $\pm$ 0.003 & 0.993 $\pm$ 0.003 & 0.992 $\pm$ 0.002 \\
 1 and 3 & 0.994 $\pm$ 0.002 & 0.994 $\pm$ 0.004 & 0.994 $\pm$ 0.003 & 0.994 $\pm$ 0.002 \\
 1 and 4 & 0.997 $\pm$ 0.001 & 0.996 $\pm$ 0.002 & 0.996 $\pm$ 0.002 & 0.996 $\pm$ 0.001 \\
 1 and 5 & 0.995 $\pm$ 0.001 & 0.995 $\pm$ 0.002 & 0.994 $\pm$ 0.003 & 0.995 $\pm$ 0.001 \\
 1 and 6 & 0.998 $\pm$ 0.002 & 0.998 $\pm$ 0.002 & 0.997 $\pm$ 0.003 & 0.997 $\pm$ 0.002 \\
 1 and 7 & 0.994 $\pm$ 0.003 & 0.994 $\pm$ 0.003 & 0.992 $\pm$ 0.004 & 0.993 $\pm$ 0.003 \\
  1 and 8 & 0.994 $\pm$ 0.003 & 0.994 $\pm$ 0.004 & 0.993 $\pm$ 0.004 & 0.993 $\pm$ 0.003 \\
1 and 9 & 0.995 $\pm$ 0.001 & 0.995 $\pm$ 0.003 & 0.995 $\pm$ 0.002 & 0.995 $\pm$ 0.001 \\
 2 and 3 & 0.976 $\pm$ 0.004 & 0.983 $\pm$ 0.004 & 0.968 $\pm$ 0.007 & 0.976 $\pm$ 0.004 \\
2 and 4 & 0.994 $\pm$ 0.002 & 0.992 $\pm$ 0.003 & 0.996 $\pm$ 0.002 & 0.994 $\pm$ 0.002 \\
  2 and 5 & 0.980 $\pm$ 0.003 & 0.979 $\pm$ 0.004 & 0.980 $\pm$ 0.005 & 0.979 $\pm$ 0.003 \\
2 and 6 & 0.988 $\pm$ 0.002 & 0.986 $\pm$ 0.005 & 0.991 $\pm$ 0.003 & 0.988 $\pm$ 0.003 \\
 2 and 7 & 0.986 $\pm$ 0.002 & 0.988 $\pm$ 0.004 & 0.985 $\pm$ 0.004 & 0.987 $\pm$ 0.002 \\
 2 and 8 & 0.980 $\pm$ 0.003 & 0.977 $\pm$ 0.004 & 0.983 $\pm$ 0.004 & 0.980 $\pm$ 0.004 \\
 2 and 9 & 0.990 $\pm$ 0.002 & 0.988 $\pm$ 0.004 & 0.992 $\pm$ 0.004 & 0.990 $\pm$ 0.003 \\
3 and 4 & 0.998 $\pm$ 0.001 & 0.997 $\pm$ 0.002 & 0.999 $\pm$ 0.001 & 0.998 $\pm$ 0.001 \\
 3 and 5 & 0.966 $\pm$ 0.003 & 0.969 $\pm$ 0.006 & 0.957 $\pm$ 0.003 & 0.963 $\pm$ 0.003 \\
 3 and 6 & 0.997 $\pm$ 0.001 & 0.997 $\pm$ 0.002 & 0.997 $\pm$ 0.002 & 0.997 $\pm$ 0.001 \\
 3 and 7 & 0.991 $\pm$ 0.002 & 0.989 $\pm$ 0.004 & 0.995 $\pm$ 0.002 & 0.992 $\pm$ 0.002 \\
3 and 8 & 0.985 $\pm$ 0.002 & 0.982 $\pm$ 0.003 & 0.988 $\pm$ 0.003 & 0.985 $\pm$ 0.002 \\
 3 and 9 & 0.990 $\pm$ 0.002 & 0.991 $\pm$ 0.003 & 0.988 $\pm$ 0.003 & 0.989 $\pm$ 0.002 \\
4 and 5 & 0.997 $\pm$ 0.001 & 0.998 $\pm$ 0.002 & 0.996 $\pm$ 0.002 & 0.997 $\pm$ 0.001 \\
 4 and 6 & 0.993 $\pm$ 0.003 & 0.994 $\pm$ 0.003 & 0.992 $\pm$ 0.003 & 0.993 $\pm$ 0.003 \\
 4 and 7 & 0.990 $\pm$ 0.003 & 0.993 $\pm$ 0.002 & 0.988 $\pm$ 0.004 & 0.990 $\pm$ 0.003 \\
 4 and 8 & 0.994 $\pm$ 0.002 & 0.997 $\pm$ 0.003 & 0.991 $\pm$ 0.004 & 0.994 $\pm$ 0.003 \\
 4 and 9 & 0.978 $\pm$ 0.004 & 0.979 $\pm$ 0.005 & 0.977 $\pm$ 0.005 & 0.978 $\pm$ 0.004 \\
 5 and 6 & 0.989 $\pm$ 0.003 & 0.988 $\pm$ 0.005 & 0.991 $\pm$ 0.003 & 0.989 $\pm$ 0.003 \\
 5 and 7 & 0.995 $\pm$ 0.001 & 0.995 $\pm$ 0.002 & 0.997 $\pm$ 0.002 & 0.996 $\pm$ 0.001 \\
  5 and 8 & 0.989 $\pm$ 0.003 & 0.992 $\pm$ 0.003 & 0.987 $\pm$ 0.005 & 0.989 $\pm$ 0.003 \\
 5 and 9 & 0.988 $\pm$ 0.003 & 0.991 $\pm$ 0.004 & 0.987 $\pm$ 0.006 & 0.989 $\pm$ 0.003 \\
 6 and 7 & 0.998 $\pm$ 0.001 & 0.998 $\pm$ 0.002 & 0.998 $\pm$ 0.001 & 0.998 $\pm$ 0.001 \\
6 and 8 & 0.988 $\pm$ 0.002 & 0.989 $\pm$ 0.004 & 0.987 $\pm$ 0.005 & 0.988 $\pm$ 0.002 \\
 6 and 9 & 0.995 $\pm$ 0.001 & 0.996 $\pm$ 0.002 & 0.995 $\pm$ 0.002 & 0.995 $\pm$ 0.001 \\
 7 and 8 & 0.995 $\pm$ 0.002 & 0.997 $\pm$ 0.003 & 0.993 $\pm$ 0.002 & 0.995 $\pm$ 0.002 \\
 7 and 9 & 0.971 $\pm$ 0.004 & 0.961 $\pm$ 0.004 & 0.980 $\pm$ 0.006 & 0.970 $\pm$ 0.005 \\
 8 and 9 & 0.985 $\pm$ 0.003 & 0.989 $\pm$ 0.003 & 0.982 $\pm$ 0.004 & 0.985 $\pm$ 0.003 \\
\hline
\end{tabular}
}\label{tab-mnist}
\end{table}

The results presented in Table~\ref{tab-mnist} demonstrate the remarkable performance of the SQQNN architecture across various MNIST class pairings, achieving near-perfect accuracy, precision, sensitivity, and F1 scores in most cases. For simpler pairings, such as digits 0 and 1, the network attained an accuracy of $0.999 \pm 0.001$, reflecting its ability to distinguish visually distinct classes with minimal error. Even for more challenging combinations, such as 2 and 3 or 7 and 9, the accuracy remained high, at $0.976 \pm 0.004$ and $0.971 \pm 0.004$, respectively, underscoring the robustness of the reduced SQQNN model. The consistently low variance across all metrics indicates stable performance. 

\section{Comparison with other Models}\label{sec-comp}

We compared the performance of the SQQNN against results reported in the literature across several real-world datasets. For the Combined Cycle Power Plant Dataset, Ref.~\cite{Cuellar2024} reported training and test MSE values of $0.048 \pm 0.0001$ and $0.051 \pm 0.0003$, respectively, based on 30 experiments. In contrast, our method achieved an MSE approximately ten times lower, indicating a substantial improvement.

For the Communities and Crime Dataset, Ref.~\cite{Kolomoytseva2021} reported a training MSE of 0.299 using classical methods. Our network achieved a significantly lower MSE, underscoring its enhanced learning capability.

In the Wisconsin Breast Cancer Dataset, Ref.~\cite{McFarthing2024} obtained a validation accuracy of $0.94 \pm 0.024$ using a bagging ensemble with a quantum asymptotically universal multi-feature approach. Our model achieved a slightly higher accuracy of $0.956 \pm 0.028$, demonstrating comparable or improved performance.

Regarding the MNIST dataset, Ref.~\cite{Mangini2020} achieved 0.98 accuracy on digits 0 and 1, while our model improved this to $0.999 \pm 0.001$. Similarly, for digits 8 and 9, Ref.~\cite{McFarthing2024} reported $0.949 \pm 0.024$, whereas our approach reached $0.985 \pm 0.003$.

\begin{table}[!h]
\caption{Comparison of the performance of the SQQNN with results reported in the literature.}
\centering 
\renewcommand{\arraystretch}{1.2}
\begin{tabular}{lllc}
\toprule
Dataset & Literature      & SQQNN  & Metric       \\ 
\midrule
Combined Cycle Power Plant     &  $0.051 \pm 0.0003$ & $\mathbf{0.0035 \pm 0.0002}$  & MSE\\
Communities and Crime          & $0.299$             & $\mathbf{0.1195 \pm 0.0252}$ & MSE\\
Wisconsin Breast Cancer        & $0.94 \pm 0.024$    & $\mathbf{0.956 \pm 0.028}$ & Accuracy \\
MNIST (0 and 1)                & $0.98$              & $\mathbf{0.999 \pm 0.001}$ & Accuracy \\
MNIST (8 and 9)                & $0.949 \pm 0.024$   & $\mathbf{0.985 \pm 0.003}$ &  Accuracy \\
\bottomrule
\end{tabular}
\label{table-results-summary}
\end{table}

Table~\ref{table-results-summary} summarizes the comparative performance of the SQQNN across both regression and classification tasks. These results demonstrate the SQQNN's ability to outperform or match state-of-the-art quantum and classical models while maintaining a compact and hardware-efficient architecture.

\section{Discussion about Scalability}\label{sec-multiple}

Although this work focuses on single-qubit networks, in this section we discuss strategies for extending our training techniques to multi-qubit architectures. Multiple independent qubits can parallelize measurements and reduce sampling overhead, although repeated circuit executions are still required to estimate expectation values. Below, we outline a more substantial generalization.

\begin{figure}[!h]
\centering 
\includegraphics[scale=0.22]{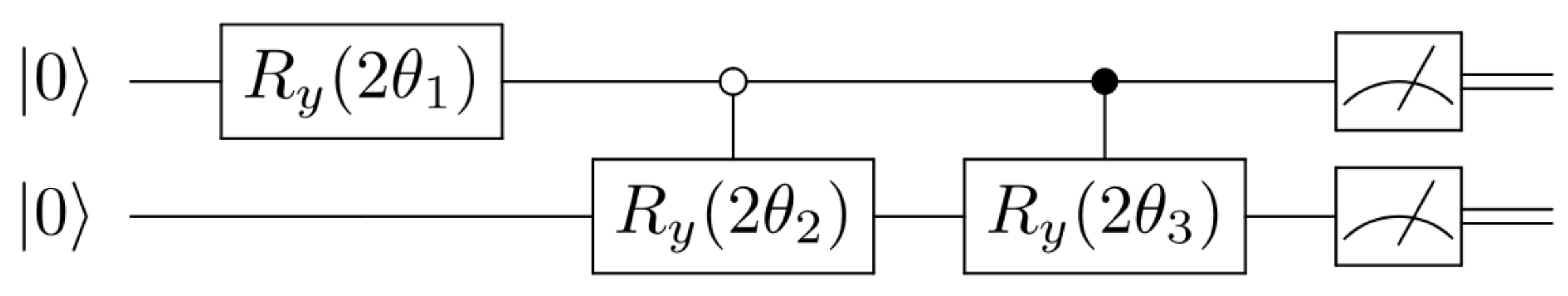}
\caption{Two-qubit network for four-class classification. This circuit is based on the standard state preparation method and can be generalized to more qubits.}
\label{fig:2-qubits}
\end{figure}

To enable multi-class classification using multiple qubits, our method can be extended via the standard state preparation approach~\cite{Schuld2021book2}. We illustrate this with a two-qubit circuit designed for four-class classification. The generalization to $d$ qubits for $2^d$-class classification follows naturally. The network structure for the two-qubit case is shown in Fig.~\ref{fig:2-qubits}.

In the 2-qubit architecture, the output of the network is the quantum state
\[
\ket{\psi} = \cos{\theta_1}\cos{\theta_2}\,\ket{00} + \cos{\theta_1}\sin{\theta_2}\,\ket{01} + \sin{\theta_1}\cos{\theta_3}\,\ket{10} + \sin{\theta_1}\sin{\theta_3}\,\ket{11},
\]
which defines a probability distribution over four computational basis states. The probability of assigning an input to class 0, 1, 2, or 3 is given by the squared amplitudes of the respective basis states: $\cos^2{\theta_1}\cos^2{\theta_2}$, $\cos^2{\theta_1}\sin^2{\theta_2}$, $\sin^2{\theta_1}\cos^2{\theta_3}$, and $\sin^2{\theta_1}\sin^2{\theta_3}$. Classification is performed by identifying the basis state with the highest probability upon measurement, where the parameters are assumed to follow the ansatz defined in Eq.~\eqref{eq-general_function}.

During training, we adopt a sequential approach in which each parameter $\theta_j$ is optimized to maximize the probability associated with a target class. For instance, to classify an input as class 3 (corresponding to $\ket{11}$), $\theta_1$ and $\theta_3$ are adjusted such that $\sin^2{\theta_1}\sin^2{\theta_3} \approx 1$, while the other components are suppressed. This is achieved by training $\theta_1$ and $\theta_3$ using a supervised binary loss function where the desired class is mapped to 1 and all others to 0. Once a parameter is determined for a specific class, the corresponding data is excluded from subsequent training steps, and the process is repeated for the remaining classes. This method generalizes naturally to $d$ qubits for $2^d$-class classification.

\section{Conclusion}\label{sec-conc}

We use a Single-Qubit Quantum Neural Network (SQQNN), a quantum machine learning framework that leverages the simplicity and efficiency of single-qubit neurons. By employing parameterized single-qubit operations, the SQQNN offers a streamlined yet powerful architecture for both regression and binary classification tasks. This design significantly reduces hardware requirements, simplifies implementation, and is compatible with near-term quantum devices, making it a practical alternative to more complex multi-qubit approaches.

We validated the SQQNN's capabilities through diverse applications. The model achieved near-zero training errors in logic gate evaluations, demonstrating its ability to handle linear and nonlinear relationships efficiently. For other regression tasks, the SQQNN successfully modeled continuous functions and real-world datasets, showcasing its adaptability and generalization. In classification tasks, increasing the software model-order setting reduced error in several classification experiments, and the model performed strongly on both synthetic and real-world datasets, including MNIST. These results highlight the SQQNN's scalability and flexibility. 

To optimize the network, we employed two learning methods: gradient descent with iterative refinement for regression tasks and a closed-form optimization approach for the transformed least-squares objective using a matrix-based dataset representation called polynomial-based linear least squares (LLS) for classification tasks. The polynomial-based LLS method offers rapid results, and its use with the reduced circuit at $K_{\mathrm{code}}=1$ eliminates the need for deep learning architectures. This demonstrates an efficient and streamlined training approach, making it a valuable contribution to both quantum and classical machine learning. This combination of training methods ensures accurate results across various tasks, highlighting the SQQNN's versatility in leveraging different optimization strategies tailored to specific problems.

The SQQNN provides a promising, efficient, and adaptable framework for practical quantum neural networks, bridging the gap between current quantum hardware and machine learning applications. Future research could explore extensions to multi-qubit systems, the development of novel activation functions, and the application of the model to diverse pattern datasets and real-world scenarios. Employing qudits for multi-ary classification, by capitalizing on the multi-output nature of quantum measurements, presents another intriguing direction.

\section*{Code Availability}
The source code used in this study is available at \url{https://github.com/leandrocarlossouza/QLM-codes/tree/main/sqqnn}.

\section*{Acknowledgements}

The authors thank Dr.~Adenilton J.~da Silva and Professor Francesco Petruccione for productive discussions.
The work of L.~C.~Souza was supported by CNPq grant number 302519/2024-6.
The work of R.~Portugal was supported by FAPERJ grant number CNE E-26/200.954/2022, and CNPq grant numbers 304645/2023-0 and 409552/2022-4.


\end{document}